\documentclass[prd,aps,preprint,tightenlines,superscriptaddress]{revtex4}
\usepackage{epsfig}
\usepackage{axodraw}
\usepackage{amsmath}
\begin{document}

%\preprint{DOE/ER/40762-274 \ UM-PP\#03-035}

\title{QCD Factorization for Spin-Dependent Cross Sections \\
in DIS and Drell-Yan Processes at Low Transverse Momentum}
\author{Xiangdong Ji}
%\email{xji@physics.umd.edu}
\affiliation{Department of Physics,
University of Maryland,
College Park, Maryland 20742, USA}
\author{Jian-Ping Ma}
%\email{majp@itp.ac.cn}
\affiliation{Institute of Theoretical Physics, Academia Sinica,
Beijing, 100080, P. R. China}\affiliation{Department of Physics,
University of Maryland, College Park, Maryland 20742, USA}
\author{Feng Yuan}
%\email{fyuan@physics.umd.edu}
\affiliation{Department of Physics,
University of Maryland,
College Park, Maryland 20742, USA}

\date{\today}
\vspace{0.5in}
\begin{abstract}
Built on a recent work on the quantum chromodynamic (QCD)
factorization for semi-inclusive deep-inelastic scattering (DIS),
we present a set of factorization formulas for the spin-dependent
DIS and Drell-Yan cross sections at low transverse momentum. The
result can be used to extract transverse-momentum dependent parton
distribution and fragmentation functions from relevant
experimental data.

\end{abstract}

\maketitle

%\section{Introduction}
{\bf 1.} Polarized hard scattering has becoming an important tool
to learn about the internal structure of hadrons. In recent years,
inclusive and semi-inclusive polarized deep-inelastic scattering
(DIS) has helped to unravel the quark helicity distributions in
the nucleon \cite{Review}. At present the polarized Relativistic
Heavy Ion Collider at Brookhaven National Laboratory is producing
data from polarized proton-proton collisions, which will provide a
direct measurement of the polarized gluon distribution
\cite{RHICSpin}.

An important class of polarized experiments involves measurement
of transverse momentum of the order $\Lambda_{\rm QCD}$. For
example, in semi-inclusive deep-inelastic scattering (SIDIS) at
moderate energy (for example, at HERMES kinematics), the tagged
final-state hadron has a transverse-momentum peaked at a few
hundred MeV. In Drell-Yan process also at moderate center-of-mass
energy, Drell-Yan pairs typically have a transverse momentum of
the same order of magnitude. Theoretical study of these processes
began with the classical work of Collins and Soper in which a
nearly back-to-back hadron pair is produced in $e^+e^-$ collisions
\cite{ColSop81}. A factorization theorem for the process was
established, which involves a new class of non-perturbative
hadronic observables depending on the transverse-momentum of
hadrons and/or partons: the transverse-momentum dependent (TMD)
fragmentation functions and parton distributions. In a previous
publication \cite{JiMaYu04}, we have extended the QCD
factorization theorem to the case of the semi-inclusive DIS. The
correction to the factorization is on the order of $P_\perp^2/Q^2$
and $M_h^2/Q^2$, where $P_\perp$ is the transverse momentum of the
produced hadron and $M_H$ is a hadron mass scale. In this paper,
we extend factorization further to a class of spin-dependent DIS
and Drell-Yan processes. Since the details are similar, here we
focus mostly on the final result, omitting technicalities which
can be found in the above references. The result can be used to
extract a new class of TMD parton functions from relevant
experimental data, just like the standard QCD factorization
theorems allow extraction of the Feynman parton distributions from
hard scattering data.

Before proceeding future, let us remark that there have been many
studies in the literature on the same processes at large, but not
too large transverse-momentum ($Q\gg {P_\perp}\gg\Lambda_{\rm
QCD}$), where $Q$ is hard-collision scale (for example, the
virtual-photon mass). In this case, the cross section can in
principle be calculated using the conventional perturbative QCD
method with the integrated Feynman parton distributions. However,
the hard part contains the large double logarithms of the type
$\alpha_s\ln^2P_{\perp}^2/Q^2$, which must be re-summed to make
reliable predictions \cite{DokDyaTro80,ParPet79}. The formalism
developed by Collins and Soper in the case of $e^+e^-$
annihilation \cite{ColSop81} and followed in our previous paper,
is ideal for making this type of re-summation. In fact, an
application to the unpolarized Drell-Yan process was first made by
Collins, Soper and Sterman (CSS) \cite{ColSopSte85}. Various
applications of CSS formalism to Drell-Yan, heavy boson
productions, and semi-inclusive DIS scattering have been developed
in the literature
\cite{DavSti84,ArnKau91,LadYua94,MenOlnSop96,NadStuYua00,QiuZha01}.

{\bf 2.} Let us first define a new class of non-perturbative
hadronic matrix elements, the spin and transverse-momentum
dependent parton distributions and fragmentation functions, which
one hopes to learn from high-energy scattering. As illustrated in
Fig.~1, we consider a hadron with momentum $P^\mu$ moving in the
$z$-direction, and is polarized with a spin vector $S^\mu$
(dimensionless and $S\cdot P=0$). In the limit $P^3\rightarrow
\infty$, the $P^\mu $ is proportional to the light-cone vector
$p^\mu=\Lambda(1,0,0,1)$, where $\Lambda$ is a mass dimension-1
parameter. The conjugation light-cone vector is
$n=(1/2\Lambda)(1,0,0-1)$, such that $p^2=n^2=0$ and $p\cdot n=1$.
We use the light-cone coordinates $k^\pm = (k^0\pm k^3)/\sqrt{2}$,
and write any four-vector $k^\mu$ in the form of $(k^-,\vec{k})=
(k^-, k^+, \vec{k}_\perp)$, where $\vec{k}_\perp$ represents two
perpendicular components $(k^x,k^y)$. Let $(xP^+, \vec{k}_\perp)$
represent the momentum of a parton (quark or gluon) in the hadron
as shown in Fig.~1. In a non-singular gauge (e.g. Feynman gauge),
the TMD parton distributions can be defined through the following
density matrix \cite{Col89,Col02},
\begin{eqnarray}
       {\cal M}^{\pm}(x, k_\perp, \mu, x\zeta,\rho)
        &=& p^+\int
        \frac{d\xi^-}{2\pi}e^{-ix\xi^-P^+}\int
        \frac{d^2\vec{b}_\perp}{(2\pi)^2} e^{i\vec{b}_\perp\cdot
        \vec{k}_\perp}         \label{tmdpd}\\
   &&    \times \frac{\left\langle PS\left|\overline{\psi}_q(\xi^-,0,\vec{b}_\perp){\cal
        L}^\dagger_{v}(\pm\infty;\xi^-,0,\vec{b}_\perp)  {\cal
        L}_{v}(\pm\infty;0)
        \psi_q(0)\right|PS\right\rangle}{S^\pm(\vec{b}_\perp, \mu^2, \rho) } \ ,
\nonumber
\end{eqnarray}
where $\psi_q$ is the quark field, and the Dirac- and color
indices of the quark fields are implicit.  The $+$($-$)
superscript is appropriate for DIS (Drell-Yan) process
\cite{Col02,BelJiYua03}. $v^\mu$ is a time-like dimensionless
($v^2>0$) four-vector with zero transverse components
$(v^-,v^+,\vec{0})$ and $v^-\gg v^+$. Thus the $v^\mu$ is a quasi
light-cone vector, approaching $n^\mu$. The variable $\zeta^2$
denotes the combination $(2P\cdot v)^2/v^2=\zeta^2$. ${\cal L}_v$
is a gauge link along $v^\mu$,
\begin{equation}
    {\cal L}_{v}(\pm\infty;\xi) = \exp\left(-ig\int^{\pm\infty}_0 d\lambda v\cdot A(\lambda
    v +\xi)\right) \ .
\end{equation}
Here the non-light-like gauge link is introduced to regulate the
light-cone singularities. We avoid the use of singular gauges
(e.g. the light-cone gauge) because in those gauges the gauge
potential may not vanish at infinity and gauge links at infinity
might be necessary to define gauge-invariant parton distributions
\cite{BelJiYua03}.

\begin{figure}
\begin{center}
\begin{picture}(100,70)(0,0)
\SetOffset(10,10) \SetWidth{0.7}
\ArrowLine(8,20)(8,56)\ArrowLine(72,56)(72,20)
\GOval(40,22)(34,15)(90){0.7}\DashLine(40,65)(40,-10){5}
\SetWidth{2.0}\ArrowLine(-6,0)(8,17) \ArrowLine(72,17)(86,0)
\Text(0,-5)[l]{$(P,S)$}\Text(6,66)[l]{$(x,\vec k_\perp)$}
%\Text(94,0)[l]{$P$}
\end{picture}
\end{center}
\caption{Transverse momentum dependent quark distributions in the
nucleon.}
\end{figure}
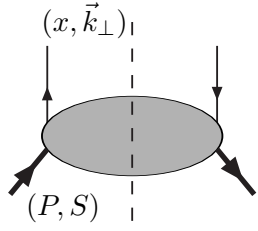

In the above definition, we have derived a soft factor defined as
\cite{JiMaYu04}:
\begin{equation}
   S^{\pm}(\vec{b}_\perp, \mu^2, \rho) =\frac{1}{N_c} \langle 0|
   {\cal L}^\dagger_{\tilde vil}( \vec{b}_\perp, -\infty)
   {\cal L}^\dagger_{vlj}(\pm\infty;\vec{b}_\perp)
   {\cal L}_{vjk}(\pm\infty;0)
    {\cal L}_{\tilde v ki}(0;-\infty) |0\rangle\ ,
\label{soft}
\end{equation}
where $i,j,k,l =1,2,3$ are color indices and new quasi light-cone
vector $\tilde v^\mu = (\tilde v^-, \tilde v^+, \vec{0})$ has been
introduced with $\tilde v^-\ll \tilde v^+$. The $\rho$ parameter
is defined as $\rho = \sqrt{v^-\tilde v^+/v^+\tilde v^-}\gg 1$.
The above soft factor will also be present in the factorization
theorems below.

The leading order expansion of the density matrix ${\cal M}$
contains eight quark distributions \cite{MulTan96,BoeMul98},
\begin{eqnarray}
{\cal M}&=&\frac{1}{2}\left[q(x,k_\perp)\not\! p+\frac{1}{M}\delta
q(x,k_\perp)\sigma^{\mu\nu}k_\mu p_\nu+\Delta q_L(x,k_\perp) \lambda \gamma_5\not\! p  \right.\nonumber\\
&& +\frac{1}{M}\Delta q_T(x,k_\perp)\gamma_5\not\!
p(\vec{k_\perp}\cdot \vec{S}_\perp)+\frac{1}{M}\delta
q_L(x,k_\perp)\lambda i\sigma_{\mu\nu}\gamma_5 p^\mu k_\perp^\nu
+\delta q_T(x, k_\perp)i\sigma_{\mu\nu}\gamma_5 p^\mu S_\perp^\nu
\nonumber\\
&&\left.+\frac{1}{M^2}\delta
q_{T'}(x,k_\perp)i\sigma_{\mu\nu}\gamma_5 p^\mu
\left(\vec{k}_\perp\cdot\vec{S}_\perp
k_\perp^\nu-\frac{1}{2}\vec{k}_\perp^2S_\perp^\nu\right)
+\frac{1}{M}q_T(x,k_\perp)\epsilon^{\mu\nu\alpha\beta}\gamma_\mu
p_\nu k_\alpha S_\beta \right]\ , \label{tmdpar}
\end{eqnarray}
where $M$ is the nucleon mass. We have omitted $\pm$ labels and
the arguments $\zeta$ and $\mu$ for the distributions at the right
side of the equation. The convention for $\gamma_5$ and
$\epsilon$-tensor follows that of \cite{ItzZub}. In principle,
there are Lorentz structures depending on $v^\mu$ and $\tilde
v^\mu$; they can taken to be $n^\mu$ and $p^\mu$, respectively,
when no light-cone singularity is present. The polarization vector
$S^\mu$ has been decomposed into a longitudinal component
$S_L^\mu$ and a transverse one $S_\perp^\mu$,
\begin{eqnarray}
   S^\mu &=& S^\mu_L + S^\mu_T \nonumber \\
    &=&  \lambda \left (  \frac{p^\mu}{M} -\frac{M}{2}n^\mu
  \right)  + S^\mu_T
\end{eqnarray}
where $\lambda$ is the helicity. The notations for the
distributions follow Ref. \cite{JiMaYu03}, which are different
than those in \cite{MulTan96,BoeMul98}.

Out of the eight TMD distributions, three of them are associated
with the $k_\perp$-even structure under the exchange $k_\perp
\rightarrow -k_\perp$: $q(x,k_\perp)$, $\Delta q_L(x,k_\perp)$,
and $\delta q_T(x,k_\perp)$, which correspond to the unpolarized,
longitudinal polarized, and transversity distributions,
respectively \cite{JafJi91}. These distributions will survive
after integrating over transverse momentum. The other five
distributions are associated with the $k_\perp$-odd structures,
and hence vanish when $k_\perp$ are integrated. Two of them,
$q_T(x,k_\perp)$ and $\delta q(x,k_\perp)$, are odd under naive
time-reversal transformation \cite{Col93} and require hadronic
final-state interactions to be non-zero \cite{BroHwaSch02}.

If one is interested in production of hadrons, TMD fragmentation
functions must be introduced. Following the above, we define the
following density matrix for a quark fragmentation into a (pseudo)
scalar hadron,
\begin{eqnarray}
  {\cal M}_h(z, P_{h\perp},\mu,\hat \zeta/z, \rho) &=&\frac{n^-}{z}
  \int \frac{d\xi^-}{2\pi}\frac{d^2 \vec{b}}{(2\pi)^2}
  e^{-i(k^+\xi^--\vec{k}_\perp\cdot\vec{b}_\perp)}
 \label{ffdef} \\
  && \times \sum_X \frac{1}{3}\sum_a\langle 0|{\cal L}_{\tilde v}(-\infty,0)\psi_{\beta a}(0)
   |P_hX\rangle \gamma^+_{\alpha\beta}
 \nonumber \\ && \times \langle P_hX|(\overline{\psi}_{\alpha a}(\xi^-,\vec{b}) {\cal
L}_{\tilde v}^\dagger(\xi^-,\vec{b},-\infty)|0\rangle/S(b_\perp,
\mu, \rho) \nonumber \ ,
\end{eqnarray}
where $\tilde v$ is mainly along the light-cone direction
conjugating to $P_h$, $k^+=P^+_h/z$ and $\vec{k}_\perp =
-\vec{P}_{h\perp}/z$, and $a=1,2,3$ is a color index. The variable
$\hat \zeta$ is defined as $ \hat \zeta^2 = 4(P_h\cdot \tilde
v)^2/\tilde v^2$. At leading order, we can have two fragmentation
functions in the expansion,
\begin{eqnarray}
{\cal M}_h&=&\frac{1}{2}\left[\hat q(x,p_\perp)\not\!
n+\frac{1}{M}\delta \hat q(x,p_\perp)\sigma^{\mu\nu}p_{\mu\perp}
n_\nu \right] \ . \label{tmdff}
\end{eqnarray}
The second one is (naively) time-reversal odd.

The $k_\perp$-even parton distributions satisfy the same
Collins-Soper evolution equation respect to $\zeta$, e.g.,
\cite{ColSop81}
\begin{equation}
    \zeta\frac{\partial}{\partial \zeta}f(x,b,\mu,x\zeta)
      = \big(K(b,\mu)+G(x\zeta,\mu)\big)f(x,b,\mu,x\zeta) \ ,
\label{zetarg}
\end{equation}
where $f=q,\Delta q_L,$ and $\delta q_T$, and the sum $K+G$ is
independent of ultraviolet scale $\mu$. This is because the
$\zeta$-dependence arises from light-cone divergences that are
independent of the spin structure. We have verified this
explicitly at one-loop order. For $k_\perp$-odd distributions,
similar equations have not yet been studied in the literature.

{\bf 3.} In \cite{JiMaYu04}, a factorization formula has been
shown for the cross section of unpolarized semi-inclusive DIS,
which involves the unpolarized TMD parton distributions and
fragmentation functions. Let us first extend this result to
polarized semi-inclusive DIS process. Our result is valid up to
power corrections in $P_\perp^2/Q^2$, and can be used to extract
the new non-perturbative hadronic matrix elements from relevant
experimental data.

The differential cross section for semi-inclusive DIS reads
\begin{equation}
    \frac{d\sigma}{dx_Bdydz_hd^2\vec{P}_{h\perp}}
      = \frac{2\pi\alpha^2_{\rm em}}{Q^4}
      y\ell_{\mu\nu}W^{\mu\nu}(P, q, P_h) \ ,
\end{equation}
where the leptonic tensor is
\begin{eqnarray}
\ell^{\mu\nu} &=& 2(\ell^\mu{\ell'}^\nu + \ell^\mu{\ell'}^\nu -
g^{\mu\nu}Q^2/2 -2i\lambda_\ell \epsilon^{\mu\nu\alpha\beta}
{\ell'}_\alpha \ell_\beta) \ ,
\end{eqnarray}
and $\lambda_\ell$ is the helicity of the initial lepton, $\ell$
and $\ell'$  the initial and final momenta of the lepton, $q^\mu
=\ell^\mu-{\ell'}^\mu$ the momentum of the virtual photon, $P_h$
the momentum of the observed hadron. As usual, $x_B$ is the
Bjorken variable, $y$ is the fraction of the lepton energy loss,
and $Q^2=-q^2$. The hadron tensor has the following expression in
QCD,
\begin{equation}
   W^{\mu\nu}(P, q, P_h) =\frac{1}{4z_h}\sum_X
   \int \frac{d^4\xi}{(2\pi)^4} e^{iq\cdot \xi}
   \langle PS|J_\mu(\xi)|XP_h\rangle \langle XP_h|J_\nu(0)|PS\rangle
   \ ,
\end{equation}
where $J^\mu$ is the electromagnetic current of the quarks, $X$
represents all other final-state hadrons other than the observed
particle $h$. The variable $z_h$ can be defined as $P\cdot
P_h/P\cdot q$ or $P_h^-/q^-$.

In a coordinate system in which the virtual-photon and hadron
momenta are collinear and along the $z$-direction, the above
hadronic tensor $W^{\mu\nu}$ has the following leading-twist
structures,
\begin{eqnarray}
2W^{\mu\nu}&=&-g^{\mu\nu}_\perp F_{UU}^{(1)}\nonumber\\
&&+i \lambda_\ell \epsilon^{\mu\nu}_\perp (\lambda
F_{LL}+S^i_\perp \hat{P}^i_{h\perp}F_{LT}) \nonumber\\
&&+(g^{\mu\nu}_\perp-\hat{P}^\mu_{h\perp}\hat{P}^\nu_{h\perp}-\hat{P}^\nu_{h\perp}\hat{P}^\mu_{h\perp})F_{UU}^{(2)}\nonumber\\
&&+\lambda\hat{P}_{h\perp\alpha}\epsilon_\perp^{\alpha(\mu}\hat{P}_{h\perp}^{\nu)} F_{UL}\nonumber\\
&&-g^{\mu\nu}_\perp\epsilon^{\alpha\beta}S_{\perp\alpha}\hat P_{h\perp\beta}F_{UT}^{(1)}\nonumber\\
&&+\left(S_{\perp\alpha}\epsilon_\perp^{\alpha(\mu}\hat{P}^{\nu)}_{h\perp}
   -g^{\mu\nu}_\perp\epsilon^{\alpha\beta}S_{\perp\alpha}\hat P_{h\perp\beta} \right)
     (F_{UT}^{(2)}-F_{UT}^{(3)}/2)\nonumber\\
&&+\hat{P}_{h\perp\alpha}\epsilon_\perp^{\alpha(\mu}\hat{P}_{h\perp}^{\nu)}(\hat{\vec{P}}_{h\perp}\cdot
\vec{S}_\perp) F_{UT}^{(3)} \ ,
\end{eqnarray}
where $g^{\mu\nu}_\perp=g^{\mu\nu}-p^\mu n^\nu-p^\nu n^\nu$ and
$\epsilon^{\mu\nu}_\perp=\epsilon^{\alpha\beta\mu\nu}p_\alpha
n_\beta$, and $\hat P_{h\perp}$ is the unit vector along
$\vec{P}_{h\perp}$. $F_{I_1I_2}$ are {\it structure functions}
depending on $x_B$, $z_h$, $P_{h\perp}$, and $Q^2$, where $I_1$
represents the polarization of the incident lepton and $I_2$ that
of the target hadron ($U$: unpolarized, $L$:
longitudinally-polarized, $T$: transversely-polarized). For
example, $F_{UU}$ denotes the unpolarized structure function,
$F_{UT}$ means the target is transversely polarized while the beam
lepton is unpolarized.

Substituting the hadronic tensor into the differential cross
section, we get
\begin{eqnarray}
    \frac{d\sigma}{dx_Bdydz_hd^2\vec {P}_{h\perp}}
      &=& \frac{4\pi\alpha^2_{\rm em}s}{Q^4}\left[(1-y+y^2/2)
      x_B (F_{UU}^{(1)}+\sin(\phi_h-\phi_S) |S_\perp| F_{UT}^{(1)})\right. \nonumber\\
      && +\lambda_\ell y(1-y/2)x_B(\lambda F_{LL}+\cos(\phi_h-\phi_S)|S_\perp|F_{LT})\nonumber\\
      &&+(1-y)x_B\left(-\cos(2\phi_h) F_{UU}^{(2)}+\lambda\sin(2\phi_h)
      F_{UL}\right.\nonumber\\
      &&\left.\left.+|\vec{S}_\perp|\sin(\phi_h+\phi_S) F_{UT}^{(2)}+\frac{1}{2}|\vec{S}_\perp|\sin(3\phi_h-\phi_S)
      F_{UT}^{(3)}\right)
      \right] \ ,
\end{eqnarray}
where $s =(P+\ell)^2$, and $\phi_{S}$ and $\phi_{h}$ are azimuthal
angles for the transverse polarization vector of the initial
hadron and the transverse momentum of the final state hadron,
respectively. By studying the angular dependence of the cross
section, one can isolate the different structure functions.

Following \cite{JiMaYu04}, we can factorize the structure
functions into TMD parton distributions and fragmentation
functions, and soft and hard parts. For example, the double
polarized structure function $F_{LL}$ has the following
factorization form,
\begin{eqnarray}
F_{LL}(x_B,z_h,Q^2,P_{h\perp})&=&\sum_{q=u,d,s,...} e_q^2\int
d^2\vec{k}_{\perp} d^2\vec {p}_{\perp}
      d^2\vec{\ell}_\perp
   \nonumber \\
   && \times  \Delta q_L\left({x_B}, k_{\perp},\mu^2,x_B\zeta, \rho\right)
    \hat q\left({z_h}, p_{\perp},\mu^2,\hat\zeta/z_h, \rho\right)
    S^+(\vec{\ell}_\perp,\mu^2,\rho) \nonumber \\
&& \times H_{LL}\left(Q^2/\mu^2,\rho\right)
\delta^2(z\vec{k}_{\perp}+\vec{p}_{\perp} +\vec{\ell}_\perp-
\vec{P}_{h\perp}) \ .
\end{eqnarray}
where we have chosen a coordinate system in which $\zeta^2 =
(Q^2/x_B^2)\rho$ and $\hat \zeta^2 = (Q^2z_h^2)\rho$. Similarly,
we can have for the other structure functions,
\begin{eqnarray}
F_{UU}^{(2)}&=&\int \frac{2\vec{k}_\perp\cdot
\hat{\vec{P}}_{h\perp}\vec{p}_\perp\cdot
\hat{\vec{P}}_{h\perp}-\vec{k}_\perp\cdot \vec{p}_\perp}{MM_h}
\delta q\left({x_B}, k_{\perp}\right)\delta \hat
q\left({z_h},p_{\perp}\right)
    S^+(\vec{\ell}_\perp)H_{UU}^{(2)}\left(Q^2\right) \ , \nonumber\\
F_{UL}&=&\int \frac{2\vec{k}_\perp\cdot
\hat{\vec{P}}_{h\perp}\vec{p}_\perp\cdot
\hat{\vec{P}}_{h\perp}-\vec{k}_\perp\cdot \vec{p}_\perp}{MM_h}
\delta q_L\left({x_B}, k_{\perp}\right)\delta \hat
q\left({z_h},p_{\perp}\right)
    S^+(\vec{\ell}_\perp)H_{UL}\left(Q^2\right) \ , \nonumber\\
F_{LT}&=&\int \frac{\vec{k}_\perp\cdot \hat{\vec{P}}_{h\perp}}{M}
\Delta q_T\left({x_B}, k_{\perp}\right)
    \hat q\left({z_h}, p_{\perp}\right)
    S^+(\vec{\ell}_\perp)  H_{LT}\left(Q^2\right) \ ,\nonumber\\
F_{UT}^{(1)}&=&\int \frac{\vec{k}_\perp\cdot
\hat{\vec{P}}_{h\perp}}{M} q_T\left({x_B}, k_{\perp}\right)\hat
q\left({z_h},p_{\perp}\right)
    S^+(\vec{\ell}_\perp)H_{UT}^{(1)}\left(Q^2\right) \ , \nonumber\\
F_{UT}^{(2)}&=&\int \frac{\vec{p}_\perp\cdot
\hat{\vec{P}}_{h\perp}}{M} \delta q_T\left({x_B},
k_{\perp}\right)\delta\hat q\left({z_h},p_{\perp}\right)
    S^+(\vec{\ell}_\perp)H_{UT}^{(2)}\left(Q^2\right) \ , \nonumber\\
F_{UT}^{(3)}&=&\int \frac{4(\vec{k}_\perp\cdot
\hat{\vec{P}}_{h\perp})^2\vec{p}_\perp\cdot
\hat{\vec{P}}_{h\perp}-|\vec{k}_\perp|^2 \vec p_\perp\cdot
\hat{\vec{P}}_{h\perp}-2\vec{k}_\perp\cdot \vec{p}_\perp
\vec{k}_\perp\cdot \hat{\vec{P}}_{h\perp}}{M^2M_h} \nonumber\\
&&\times \delta q_T'\left({x_B}, k_{\perp}\right)\delta \hat
q\left({z_h},p_{\perp}\right)
    S^+(\vec{\ell}_\perp)H_{UT}^{(3)}\left(Q^2\right) \ ,
\end{eqnarray}
where the simple integral symbol represents a complicated integral
used above, i.e., $\int=\int d^2\vec{k}_{\perp} d^2\vec
{p}_{\perp}d^2\vec{\ell}_\perp\delta^2(z\vec{k}_{\perp}+\vec{p}_{\perp}
+\vec{\ell}_\perp- \vec{P}_{h\perp})$. And for simplicity, we also
omit the explicit sum over all flavors weighted with their charge
square, and the explicit $\mu$, $\rho$ and $\zeta$ dependence for
the parton distributions and fragmentation functions, as well as
the soft and hard factors.

From the above factorization formulas in the transverse momentum
space, we can also obtain the factorization form in the impact
parameter space. For example,
\begin{eqnarray}
   \tilde F_{UT}^{(1)}(b)&=& \frac{-1}{Mz_h}\partial^i_b\left[
      \left(\partial^i_b q_T(x_B, z_hb)\right)\hat q\left(z_h,b\right)
 S^+(b) H_{UT}^{(1)}\left(Q^2\right)\right]\ ,
\end{eqnarray}
where $\tilde F_{UT}^{(1)}(b)=\int
d^2P_{h\perp}e^{i\vec{P}_{h\perp}\cdot
\vec{b}}|\vec{P}_{h\perp}|F_{UT}^{(1)}(P_{h\perp})$, and other
quantities depending on $\vec{b}$ are obtained from the Fourier
transformation of the corresponding momentum-dependent ones. The
convolution in transverse momentum space becomes a simple product
in the impact parameter space, while the $\vec{k}_\perp$ moment
integral generates a derivative on $\vec{b}$.

An explicit one-loop calculation for $F_{UU}$ and $F_{LL}$ can be
done.  The hard parts at one-loop order have the following form,
\begin{eqnarray}
H_{UU}^{(1)}&=&H_{LL}
=\frac{\alpha_s}{2\pi}C_F\left[\left(1+\ln\rho^2\right)
     \ln\frac{Q^2}{\mu^2}
     - \ln\rho^2 + \frac{1}{4}\ln^2\rho^2 + \pi^2-4\right] \ .
\end{eqnarray}
To get the next-to-leading order corrections for the hard parts
related to $k_\perp$-odd distributions, one has to consider
two-loop calculations for scattering off an elementary quark
target.

In the above discussions, we consider the factorization of the
leading contributions in the expansion of $P_{h\perp}^2/Q^2$.
However, the power corrections in the semi-inclusive DIS
\cite{MulTan96} might have sizable effects at moderate $Q^2$
range. They are not included in our factorization formalism. On
the other hand, it will be interested to extend our factorization
formalism to include the sub-leading power contributions in SIDIS,
just like what have been done for the sub-leading power correction
to the inclusive DIS and DY cross sections \cite{QiuSte91}. This
however is beyond the scope of the present paper.

One can apply the above formalism to study the polarized cross
sections and asymmetries in the SIDIS, and compare theoretical
predictions with experimental measurements if the non-perturbative
parton functions are known from solving non-perturbative QCD. In
practice, we can treat them as unknown inputs, and fit to the
experimental data. The parton functions determined
phenomenologically can be used to make predictions for similar
processes. Note that in the TMD quark distributions and
fragmentation functions there is a double logarithmic dependence
on the hadron energy (in terms of $\ln^2 b^2\zeta^2$) which is
controlled by the Collins-Soper evolution equations as shown in
Eq.~(\ref{zetarg}). To get the reliable prediction for the cross
sections for the SIDIS, one has to solve these evolution equations
to re-sum the double logarithms. As we showed in the above the
Collins-Soper evolutions for the $k_\perp$-even TMD quark
distributions are the same. It will be interested to study the
Collins-Soper evolution for all the leading-twist quark
distributions including the $k_\perp$-odd ones. We leave this
study in a separate publication.

Let us finally remark on the two special cases related to the
above results. The first case concerns cross sections integrated
over the transverse momentum of the hadrons \cite{MulTan96}. Since
the above factorization formulas are valid only when
$P_{h\perp}\ll Q$, one cannot use them to integrate out all
$P_{h\perp}$. To do that, one must have factorization formulas
valid at all $P_{h\perp}$, including all power corrections.
Alternatively, one can prove the factorization theorems in the
$P_{h\perp}$-integrated form, which can be done, but is beyond the
scope of this paper.

The second case concerns the region of transverse momentum
$\Lambda_{\rm QCD}\ll P_{h\perp}\ll Q$. In this region, the above
factorization formula is still valid. However, because
$P_{h\perp}$ is now hard, all structure functions can be further
factorized in terms of ordinary twist-two and twist-three parton
distributions. Using Collins and Soper equations, one can sum over
large logarithms of type $\alpha_s \ln^2 Q^2/P_{h\perp}^2$. This
case, has been studied for example in
\cite{MenOlnSop96,NadStuYua00}. Some of the relevant leading-order
perturbative QCD calculations have been done in \cite{KoiNag03}.

{\bf 4.} In the Drell-Yan processes, we have two hadrons in the
initial states: $A$ and $B$. We consider the hadron $A$ moving
along $z$-direction while $B$ along -$z$-direction. For the TMD
parton distributions in hadron $B$, we introduce another
non-light-like vector $\tilde v=(\tilde v^-,\tilde v^+,\vec{0})$
($\tilde v^-\ll \tilde v^+$) and the variable
$\overline\zeta^2=(2P_B\cdot \tilde v)^2/\tilde v^2$, just like
for the fragmentation functions in DIS. The TMD quark and
anti-quark distributions for hadron $B$ are similar to those of
hadron $A$.

We study the the production of a lepton pair with low transverse
momentum in Drell-Yan process:
\begin{equation}
A(P_1, S_1) +B (P_2,S_2) \to \gamma^* (q) +X \to \ell^+ + \ell^ -
+X,
\end{equation}
where $P_1$ and $P_2$, $S_1$ and $S_2$ are momenta and
polarization vectors of $A$ and $B$, respectively. The
differential cross sections for $\ell^+\ell^-$ production reads
\begin{eqnarray}
\frac{d\sigma}{d^4Qd\Omega}=\frac{\alpha^2_{\rm em}}{2 s
Q^4}L_{\mu\nu}W^{\mu\nu} \ ,
\end{eqnarray}
where $L_{\mu\nu}$ is the leptonic tensor. For the Drell-Yan
production, we have $L^{\mu\nu}=4(\ell_1^\mu \ell_2^\nu +
\ell_1^\mu \ell_2^\nu - g^{\mu\nu}Q^2/2)$, where $\ell_i$ is the
lepton's momentum. $\Omega$ is the solid angle of the lepton in
the virtual photon rest frame. $Q^\mu$ is the momentum of the
lepton pair; $s$ is the total center-of-mass energy square. The
hadronic tensor is defined as:
\begin{equation}
   W_{\mu\nu}(x_1,x_2,Q_\perp,S_1,S_2) =
   \int \frac{d^4\xi}{(2\pi)^4} e^{-iq\cdot \xi}
   \langle P_1S_1P_2S_2 |J_\mu (\xi) J_\nu(0)| P_1S_1P_2S_2 \rangle
   \ ,
\end{equation}
where $J_\mu$ is the electromagnetic current. Disregarding the
contributions from $k_\perp$-odd parton distributions, the leading
hadronic tensor can be decomposed into three terms,
\begin{equation}
2W^{\mu\nu}=-g_\perp^{\mu\nu}(W_{0}-\lambda_1\lambda_2W_{LL})-g_{TT}^{\mu\nu}
W_{TT} \ ,
\end{equation}
where $g_{TT}^{\mu\nu}=\vec{S}_{1\perp}\cdot\vec{S}_{1\perp}
g_\perp^{\mu\nu}+S_1^\mu S_2^\nu+S_1^\nu S_2^\mu$. The tensor
structure $W_{0}$ denotes the unpolarized tensor, $W_{LL}$ the
double longitudinal polarized tensor, and $W_{TT}$ the double
transversely polarized tensor. We can choose a frame that the
momenta $P_1^\mu$, $P_2^\mu$ and $Q^\mu$ can be decomposed into
the following forms,
\begin{eqnarray}
P_1^\mu&=&p^\mu+\frac{M^2}{2}n^\mu \ , \nonumber\\
P_2^\mu&=&\frac{s}{2}n^\mu+\frac{M^2}{s}p^\mu \ , \nonumber\\
Q^\mu&=&x_1p^\mu+x_2\frac{s}{2}n^\mu+Q_\perp^\mu\ .
\end{eqnarray}
The polarization vectors of hadrons  can be decomposed as:
\begin{eqnarray}
S_1^\mu &=& \lambda_1 \left ( \frac{p^\mu}{M}
-\frac{M}{2} n^\mu \right) + S_{1\perp}^\mu  \ , \nonumber\\
S_2^\mu &=& \lambda_2 \left ( \frac{s}{2M} n^\mu -\frac{M}{s}
p^\mu \right ) + S_{2\perp}^\mu \ ,
\end{eqnarray}
where $\lambda_1$ and $\lambda_2$  are the helicities of hadron
$A$ and $B$ respectively, and $S_{i\perp}$ the relevant transverse
polarization vectors.

Using the hadronic tensor, the differential cross section can be
expressed as
\begin{eqnarray}
\frac{d\sigma}{d^4Qd\Omega}&=&\frac{\alpha^2}{2sQ^2}\left\{(1+\cos^2\theta)W_{0}(x_1,x_2,Q^2,Q_\perp)\right.\nonumber\\
    &&-(1+\cos^2\theta)\lambda_1\lambda_2W_{LL}(x_1,x_2,Q^2,Q_\perp)\nonumber\\
    &&\left.+\sin^2\theta\cos(\phi_1+\phi_2)|\vec{S}_{1\perp}||\vec{S}_{2\perp}|W_{TT}(x_1,x_2,Q^2,Q_\perp)\right\}\
    .
\end{eqnarray}
where $\theta$ is the polar angle of the lepton momentum, and
$\phi_1$ and $\phi_2$ are azimuthal angles of the nucleon
transverse spin relative to the lepton plane. $W_{LL}$ contributes
to double longitudinal spin asymmetry and $W_{TT}$ to double
transverse spin asymmetries.

Following Ref.~\cite{JiMaYu04}, we have the following
factorization formulas for the above hadronic tensors,
\begin{eqnarray}
 W_{0}(x_1,x_2,Q^2,Q_\perp)&=&\sum_{q=u,d,s,...} \frac{e_q^2}{3}\int d^2\vec{k}_{1\perp} d^2\vec {k}_{2\perp}
      d^2\vec{\ell}_\perp
   \nonumber \\
   && \times  q\left({x_1}, k_{1\perp},\mu^2,x_1\zeta, \rho\right)
    \overline q\left({x_2}, k_{2\perp},\mu^2,x_2\overline\zeta, \rho\right)
    S^-(\vec{\ell}_\perp,\mu^2,\rho) \nonumber \\
&& \times H_0\left(Q^2,\mu^2,\rho\right)
\delta^2(\vec{k}_{1\perp}+\vec{k}_{2\perp} +\vec{\ell}_\perp-
\vec{Q}_{\perp}) \ ,\label{e1}\\
 W_{LL}(x_1,x_2,Q^2,Q_\perp)&=&\sum_{q=u,d,s,...} \frac{e_q^2}{3}\int d^2\vec{k}_{1\perp} d^2\vec {k}_{2\perp}
      d^2\vec{\ell}_\perp
   \nonumber \\
   && \times  \Delta q_L\left({x_1}, k_{1\perp},\mu^2,x_1\zeta, \rho\right)
    \Delta \overline q_L\left({x_2}, k_{2\perp},\mu^2,x_2\overline\zeta, \rho\right)
    S^-(\vec{\ell}_\perp,\mu^2,\rho) \nonumber \\
&& \times H_{LL}\left(Q^2,\mu^2,\rho\right)
\delta^2(\vec{k}_{1\perp}+\vec{k}_{2\perp} +\vec{\ell}_\perp-
\vec{Q}_{\perp}) \ ,\label{e2}\\
 W_{TT}(x_1,x_2,Q^2,Q_\perp)&=&\sum_{q=u,d,s,...} \frac{e_q^2}{3}\int d^2\vec{k}_{1\perp} d^2\vec {k}_{2\perp}
      d^2\vec{\ell}_\perp
   \nonumber \\
   && \times  \delta q_T\left({x_1}, k_{1\perp},\mu^2,x_1\zeta, \rho\right)
    \delta \overline q_T\left({x_2}, k_{2\perp},\mu^2,x_2\overline\zeta, \rho\right)
    S^-(\vec{\ell}_\perp,\mu^2,\rho) \nonumber \\
&& \times H_{TT}\left(Q^2,\mu^2,\rho\right)
\delta^2(\vec{k}_{1\perp}+\vec{k}_{2\perp} +\vec{\ell}_\perp-
\vec{Q}_{\perp}) \ ,\label{e3}
\end{eqnarray}
where we have chosen a special coordinate system: $\zeta^2 =
(Q^2/x_1^2)\rho$ and $\overline \zeta^2 = (Q^2/x_2^2)\rho$ and
$\rho=\sqrt{v^-\tilde v^+/v^+\tilde v^-}$. The above results are
accurate at leading powers in $(Q_{\perp}^2/Q^2)$ for soft
$Q_{\perp}\sim \Lambda_{\rm QCD}$. Introducing the
impact-parameter space representation, the above factorization
formulas become simple products, e.g., for the unpolarized tensor,
\begin{eqnarray}
   W_{0}(x_1,x_2,b, Q^2) &=& \sum_{q=u,d,s,...} \frac{e_q^2}{3}
      q\left(x_1, b, \mu^2, x_1\zeta, \rho\right)\overline q\left(x_2,b,\mu^2,
   x_2\overline \zeta, \rho\right) \nonumber \\
   && \times S^-(b,\mu^2, \rho) H_0\left(Q^2,\mu^2,\rho \right)\ .
\end{eqnarray}
The other two hadronic tensors can be obtained similarly.

In the following, we will show that the above factorization
formulas are valid at one-loop order, and the relevant hard parts
emerge from the calculation. We first consider the soft factor in
the factorization formulas. Since it has no spin dependence, it
contributes equally to all three structure functions. From its
definition Eq.~(\ref{soft}), the one-loop contribution Drell-Yan
soft factor is,
\begin{equation}
S^{-}(\vec{\ell}_\perp,\mu^2,\rho)=\frac{\alpha_sC_F}{2\pi^2}
  \left[\ln\frac{4(v\cdot \tilde v)^2}{v^2\tilde
    v^2}-2\right]\cdot \left[
     \frac{1}{\ell_\perp^2+\lambda^2}-\pi \delta^2(\ell_\perp)\ln \frac{\mu^2}{\lambda^2}
      \right]\ .\label{softdy}
\end{equation}
At this order, it is the same as that for the DIS process
\cite{JiMaYu04}.

\begin{figure}
\SetScale{0.8}
\begin{center} \begin{picture}(340,200)(0,0)
 \SetWidth{1.1}

\SetOffset(0,100) \ArrowLine(0,0)(20,50)\ArrowLine(20,50)(0,100)
\Photon(20,50)(40,50){2}{4}\Photon(60,50)(80,50){2}{4}\ArrowArc(50,50)(10,0,180)\ArrowArc(50,50)(10,180,360)\DashLine(50,0)(50,100){2}
\ArrowLine(80,50)(100,0)\ArrowLine(100,100)(80,50)
\GlueArc(10,25)(10,-113,67){2.0}{5}

\SetOffset(120,100)
\ArrowLine(0,0)(20,50)\ArrowLine(20,50)(0,100)\Photon(20,50)(40,50){2}{4}\Photon(60,50)(80,50){2}{4}\ArrowArc(50,50)(10,0,180)\ArrowArc(50,50)(10,180,360)\DashLine(50,0)(50,100){2}
\ArrowLine(80,50)(100,0)\ArrowLine(100,100)(80,50)
\GlueArc(10,75)(10,-67,113){2.0}{5}

\SetOffset(240,100)
\ArrowLine(0,0)(10,25)\ArrowLine(10,75)(0,100)\Photon(20,50)(40,50){2}{4}\Photon(60,50)(80,50){2}{4}\ArrowArc(50,50)(10,0,180)\ArrowArc(50,50)(10,180,360)\DashLine(50,0)(50,100){2}
\Line(0,0)(20,50)\Line(20,50)(0,100)\ArrowLine(80,50)(100,0)\ArrowLine(100,100)(80,50)
\Gluon(10,25)(10,75){2}{6}

\SetOffset(0,0)
\ArrowLine(0,0)(10,25)\Photon(20,50)(40,50){2}{4}\Photon(60,50)(80,50){2}{4}\ArrowArc(50,50)(10,0,180)\ArrowArc(50,50)(10,180,360)\DashLine(50,0)(50,100){2}
\Line(0,0)(20,50)\ArrowLine(20,50)(0,100)\ArrowLine(90,25)(100,0)\Line(80,50)(100,0)\ArrowLine(100,100)(80,50)
\Gluon(10,25)(90,25){2}{12}

\SetOffset(120,0)
\ArrowLine(10,75)(0,100)\Photon(20,50)(40,50){2}{4}\Photon(60,50)(80,50){2}{4}\ArrowArc(50,50)(10,0,180)\ArrowArc(50,50)(10,180,360)\DashLine(50,0)(50,100){2}
\ArrowLine(0,0)(20,50)\Line(20,50)(0,100)\ArrowLine(80,50)(100,0)\ArrowLine(100,100)(90,75)\Line(100,100)(80,50)
\Gluon(10,75)(90,75){2}{12}

\SetOffset(240,0)
\ArrowLine(0,0)(10,25)\Photon(20,50)(40,50){2}{4}\Photon(60,50)(80,50){2}{4}\ArrowArc(50,50)(10,0,180)\ArrowArc(50,50)(10,180,360)\DashLine(50,0)(50,100){2}
\Line(0,0)(20,50)\ArrowLine(20,50)(0,100)\ArrowLine(80,50)(100,0)\ArrowLine(100,100)(90,75)\Line(100,100)(80,50)
\Gluon(10,25)(90,75){2}{15}

\end{picture}
\end{center}
\caption{One-loop corrections to DY processes.}
\end{figure}
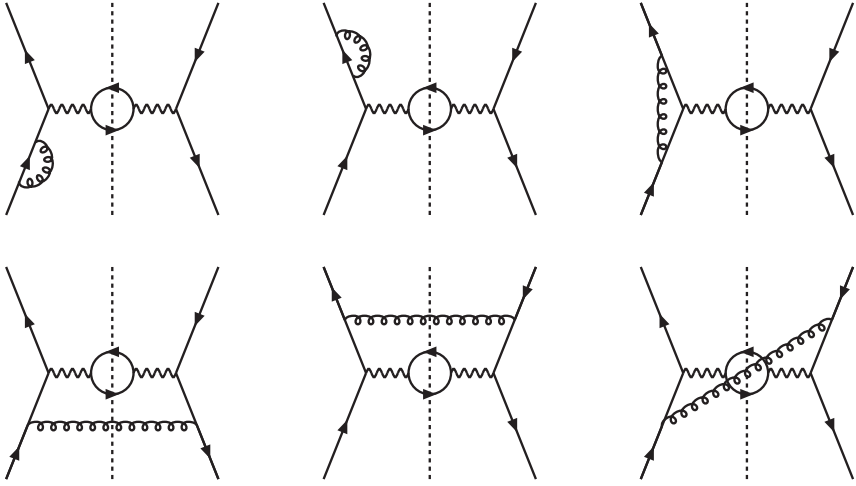

One-loop Feynman diagrams for the Drell-Yan process initiated from
a quark and anti-quark pair are shown in Fig.~2, which contain
both virtual and real corrections. The virtual corrections are the
same for all three structure functions,
\begin{equation}
    W_{0,LL,TT} = \frac{1}{3}\delta(x_1-1)\delta(x_2-1)\delta^2(\vec{Q}_{\perp})
     (1+2(Z_F-1)+2(\hat Z_V'-1))\ ,
\end{equation}
where $Z_F$ and $Z_V$ come from self-energy and vertex corrections
respectively. $Z_F$ is the same as that in DIS \cite{JiMaYu04}.
The vertex correction is different,
\begin{equation}
    \hat Z_V' = 1
    -\frac{\alpha_s}{4\pi}C_F\left(\ln\frac{Q^2}{\mu^2}
   + \ln^2\frac{Q^2}{m^2} +
    2\ln\frac{m^2}{\lambda^2}\ln\frac{Q^2}{m^2}
    -4\ln\frac{Q^2}{m^2}-\frac{4\pi^2}{3}\right) \ .
\end{equation}
which can be obtained through analytical continuation from
space-like region ($q^2<0$) to time-like region ($q^2>0$). From
the above, the one-loop TMD parton distributions, and the soft
factor in Eq.~(\ref{softdy}), the one-loop contribution to the
hard parts can be extracted.
\begin{equation}
    H_{0,LL,TT}(Q^2,\mu^2,\rho) = \frac{\alpha_s}{2\pi}C_F\left[\left(1+\ln\rho^2\right)
     \ln\frac{Q^2}{\mu^2}
     - \ln\rho^2 + \frac{1}{4}\ln^2\rho^2 + 2\pi^2-4\right] \ ,
\end{equation}
where we have chosen a coordinate system in which $x_1\zeta = x_2
\overline\zeta$ and therefore the dependence on the
quasi-light-like vectors $v$ and $\tilde v$ is simply through a
combination $\rho = \sqrt{v^-\tilde v^+/v^+\tilde v^-}$. Note that
the hard parts at this order are the same for all three structure
functions and hence are spin-independent.

The real contributions are different for the three hadronic
tensors:
\begin{eqnarray}
W_{0} &=& \frac{\alpha_sC_F}{6\pi^2}\delta(1-x_2)
     \Big \{   \frac{1-x_1}{q_\perp^2 +x_1\lambda^2+m^2 (1-x_1)^2 }
                - \frac{2 x_1(1-x_1) m^2}{(q_\perp^2 +x_1\lambda^2+m^2 (1-x_1)^2)^2}
\nonumber\\
   &&      +\frac{2x_1 }{(1-x_1)_+} \frac{1}{q_\perp^2 +x_1\lambda^2+m^2 (1-x_1)^2 }
        + \delta(1-x_1) \frac{1}{q_\perp^2+\lambda^2} \ln\frac{Q^2}{q_\perp^2+\lambda^2}
                                   \Big \}
     \nonumber\\
     &&   + (x_1\leftrightarrow x_2) \ ,
\end{eqnarray}
\begin{eqnarray}
W_{LL} &=& \frac{\alpha_sC_F}{6\pi^2}\delta(1-x_2)
        \Big \{  \frac{1-x_1}{q_\perp^2+x_1\lambda^2 +m^2 (1-x_1)^2 }
                 -\frac{2 m^2(1-x_1)(1-x_1+x_1^2)}{(q_\perp^2+x_1\lambda^2 +m^2 (1-x_1)^2)^2}
\nonumber\\
   && +
     \frac{2x_1}{(1-x_1)_+} \frac{1}{q_\perp^2 +x_1\lambda^2+m^2 (1-x_1)^2 }
    + \delta(1-x_1) \frac{1}{q_\perp^2+\lambda^2} \ln\frac{Q^2}{q_\perp^2+\lambda^2}
                                   \Big \}
        \nonumber\\
     &&   +(x_1\leftrightarrow x_2) \ ,\\
W_{TT} &=& \frac{\alpha_sC_F}{6\pi^2}\delta (1-x_2)
       \Big \{  \frac{2x_1}{(1-x_1)_+} \frac{1}{q_\perp^2 +x_1\lambda^2+m^2(1-x_1)^2}
\nonumber\\
      && - \frac{2 x_1(1-x_1) m^2}{(q_\perp^2 +x_1\lambda^2+m^2(1-x_1)^2)^2}
                 + \delta(1-x_1)\frac{1}{q_\perp^2+\lambda^2} \ln\frac{Q^2}{q_\perp^2+\lambda^2}\Big \}
      \nonumber\\
    &&  +(x_1\leftrightarrow x_2) \ .
\end{eqnarray}
All these results can be reproduced by one-loop parton
distributions for the quark and anti-quark and soft factors
discussed before. Therefore, the factorization formulas for the
Drell-Yan processes at low transverse momentum are correct at
one-loop order. For an all-order proof, one can follow the same
procedure outlined in \cite{JiMaYu04}.

Finally, we comment that when the transverse-momentum of the
lepton pair is large compared to $\Lambda_{\rm QCD}$, but much
smaller than $Q$, the above formalism leads to a resummation of
double logarithms as in the original paper by Collins, Soper, and
Sterman \cite{ColSopSte85}. For spin-dependent part, some studies
along this direction can be found in Refs. \cite{Web92,NadYua03}.

X. J. and F. Y. were supported by the U. S. Department of Energy
via grant DE-FG02-93ER-40762. J.P.M. was supported by National
Natural Science Foundation of P.R. China through grand
No.19925520. X. J. is also supported by an Overseas Outstanding
Young Chinese Scientist grant from NSFC.


\begin{thebibliography}
\frenchspacing

%\cite{Anselmino:1994gn}
%\bibitem{Anselmino:1994gn}
\bibitem{Review}
%\cite{Filippone:2001ux}
%\bibitem{Filippone:2001ux}
B.~W.~Filippone and X.~D.~Ji,
%``The spin structure of the nucleon,''
Adv.\ Nucl.\ Phys.\  {\bf 26}, 1 (2001) [arXiv:hep-ph/0101224].
%%CITATION = HEP-PH 0101224;%%
M.~Anselmino, A.~Efremov and E.~Leader,
%``The theory and phenomenology of polarized deep inelastic scattering,''
Phys.\ Rept.\  {\bf 261}, 1 (1995) [Erratum-ibid.\  {\bf 281}, 399
(1997)] [arXiv:hep-ph/9501369].
%%CITATION = HEP-PH 9501369;%%
%\cite{Barone:2001sp}
%\bibitem{Barone:2001sp}
V.~Barone, A.~Drago and P.~G.~Ratcliffe,
%``Transverse polarisation of quarks in hadrons,''
Phys.\ Rept.\  {\bf 359}, 1 (2002) [arXiv:hep-ph/0104283].
%%CITATION = HEP-PH 0104283;%%

%\cite{Bunce:2000uv}
%\bibitem{Bunce:2000uv}
\bibitem{RHICSpin}
G.~Bunce, N.~Saito, J.~Soffer and W.~Vogelsang,
%``Prospects for spin physics at RHIC,''
Ann.\ Rev.\ Nucl.\ Part.\ Sci.\  {\bf 50}, 525 (2000)
[arXiv:hep-ph/0007218].
%%CITATION = HEP-PH 0007218;%%


\bibitem{ColSop81}
%\cite{Collins:1981uk}
%\bibitem{Collins:1981uk}
J.~C.~Collins and D.~E.~Soper,
%``Back-To-Back Jets In QCD,''
Nucl.\ Phys.\ B {\bf 193}, 381 (1981) [Erratum-ibid.\ B {\bf 213},
545 (1983)];
%%CITATION = NUPHA,B193,381;%%
%\cite{Collins:va}
%\bibitem{Collins:va}
%J.~C.~Collins and D.~E.~Soper,
%``Back-To-Back Jets: Fourier Transform From B To K-Transverse,''
Nucl.\ Phys.\ B {\bf 197}, 446 (1982).
%%CITATION = NUPHA,B197,446;%%



\bibitem{JiMaYu04}
%\cite{Ji:2004wu}
%\bibitem{Ji:2004wu}
X.~Ji, J.~P.~Ma and F.~Yuan,
%``QCD factorization for semi-inclusive deep-inelastic scattering at low
%transverse momentum,''
arXiv:hep-ph/0404183.
%%CITATION = HEP-PH 0404183;%%


\bibitem{DokDyaTro80}
%\cite{Dokshitzer:dr}
%\bibitem{Dokshitzer:dr}
Y.~L.~Dokshitzer, D.~Diakonov and S.~I.~Troian,
%``Hard Semiinclusive Processes In QCD,''
Phys.\ Lett.\ B {\bf 78}, 290 (1978);
%%CITATION = PHLTA,B78,290;%%
%\cite{Dokshitzer:yd}
%\bibitem{Dokshitzer:yd}
%Y.~L.~Dokshitzer, D.~Diakonov and S.~I.~Troian,
%``On The Transverse Momentum Distribution Of Massive Lepton Pairs,''
Phys.\ Lett.\ B {\bf 79}, 269 (1978);
%%CITATION = PHLTA,B79,269;%%
%\cite{Dokshitzer:hw}
%\bibitem{Dokshitzer:hw}
%Y.~L.~Dokshitzer, D.~Diakonov and S.~I.~Troian,
%``Hard Processes In Quantum Chromodynamics,''
Phys.\ Rept.\  {\bf 58}, 269 (1980).
%%CITATION = PRPLC,58,269;%%

%\cite{Parisi:1979se}
\bibitem{ParPet79}
G.~Parisi and R.~Petronzio,
%``Small Transverse Momentum Distributions In Hard Processes,''
Nucl.\ Phys.\ B {\bf 154}, 427 (1979).
%%CITATION = NUPHA,B154,427;%%


\bibitem{ColSopSte85}
%\cite{Collins:1984kg}
%\bibitem{Collins:1984kg}
J.~C.~Collins, D.~E.~Soper and G.~Sterman,
%``Transverse Momentum Distribution In Drell-Yan Pair And W And Z Boson
%Production,''
Nucl.\ Phys.\ B {\bf 250}, 199 (1985).
%%CITATION = NUPHA,B250,199;%%


%\cite{Davies:1984hs}
\bibitem{DavSti84}
C.~T.~H.~Davies and W.~J.~Stirling,
%``Nonleading Corrections To The Drell-Yan Cross-Section At Small Transverse
%Momentum,''
Nucl.\ Phys.\ B {\bf 244}, 337 (1984);
%%CITATION = NUPHA,B244,337;%%
%\cite{Davies:1984sp}
%\bibitem{Davies:1984sp}
C.~T.~H.~Davies, B.~R.~Webber and W.~J.~Stirling,
%``Drell-Yan Cross-Sections At Small Transverse Momentum,''
Nucl.\ Phys.\ B {\bf 256}, 413 (1985).
%%CITATION = NUPHA,B256,413;%%

%\cite{Arnold:1990yk}
\bibitem{ArnKau91}
P.~B.~Arnold and R.~P.~Kauffman,
%``W And Z Production At Next-To-Leading Order: From Large Q(T) To Small,''
Nucl.\ Phys.\ B {\bf 349}, 381 (1991).
%%CITATION = NUPHA,B349,381;%%

%\cite{Ladinsky:1993zn}
\bibitem{LadYua94}
G.~A.~Ladinsky and C.~P.~Yuan,
%``The Nonperturbative regime in QCD resummation for gauge boson production at
%hadron colliders,''
Phys.\ Rev.\ D {\bf 50}, 4239 (1994) [arXiv:hep-ph/9311341].
%%CITATION = HEP-PH 9311341;%%

%\cite{Meng:1995yn}
\bibitem{MenOlnSop96}
R.~Meng, F.~I.~Olness and D.~E.~Soper,
%``Semi-Inclusive Deeply Inelastic Scattering at Small q_T,''
Phys.\ Rev.\ D {\bf 54}, 1919 (1996) [arXiv:hep-ph/9511311].
%%CITATION = HEP-PH 9511311;%%

%\cite{Nadolsky:1999kb}
\bibitem{NadStuYua00}
P.~Nadolsky, D.~R.~Stump and C.~P.~Yuan,
%``Semi-inclusive hadron production at HERA: The effect of {QCD} gluon
%resummation,''
Phys.\ Rev.\ D {\bf 61}, 014003 (2000) [Erratum-ibid.\ D {\bf 64},
059903 (2001)] [arXiv:hep-ph/9906280].
%%CITATION = HEP-PH 9906280;%%


%\cite{Qiu:2000ga}
\bibitem{QiuZha01}
J.~W.~Qiu and X.~F.~Zhang,
%``QCD prediction for heavy boson transverse momentum distributions,''
Phys.\ Rev.\ Lett.\  {\bf 86}, 2724 (2001) [arXiv:hep-ph/0012058];
%%CITATION = HEP-PH 0012058;%%
%\cite{Qiu:2000hf}
%\bibitem{Qiu:2000hf}
%J.~w.~Qiu and X.~f.~Zhang,
%``Role of the nonperturbative input in QCD resummed Drell-Yan
%Q(T)-distributions,''
Phys.\ Rev.\ D {\bf 63}, 114011 (2001) [arXiv:hep-ph/0012348].
%%CITATION = HEP-PH 0012348;%%


\bibitem{Col89}
%\cite{Collins:bt}
%\bibitem{Collins:bt}
J.~C.~Collins,
%``Sudakov Form Factors,''
Adv.\ Ser.\ Direct.\ High Energy Phys.\  {\bf 5}, 573 (1989)
[arXiv:hep-ph/0312336]; in {\it Perturbative QCD} (A.H. Mueller,
ed.) (World Scientific Publ., 1989).
%%CITATION = HEP-PH 0312336;%%

%\cite{Collins:2002kn}
\bibitem{Col02}
J.~C.~Collins,
%``Leading-twist single-transverse-spin asymmetries: Drell-Yan and
%deep-inelastic scattering,''
Phys.\ Lett.\ B {\bf 536}, 43 (2002) [arXiv:hep-ph/0204004].
%%CITATION = HEP-PH 0204004;%%

\bibitem{BelJiYua03}
%\cite{Belitsky:2002sm}
%\bibitem{Belitsky:2002sm}
A.~V.~Belitsky, X.~Ji and F.~Yuan,
%``Final state interactions and gauge invariant parton distributions,''
Nucl.\ Phys.\ B {\bf 656}, 165 (2003) [arXiv:hep-ph/0208038];
%%CITATION = HEP-PH 0208038;%%
%\cite{Ji:2002aa}
%\bibitem{Ji:2002aa}
X.~Ji and F.~Yuan,
%``Parton distributions in light-cone gauge: Where are the final-state
%interactions?,''
Phys.\ Lett.\ B {\bf 543}, 66 (2002) [arXiv:hep-ph/0206057].
%%CITATION = HEP-PH 0206057;%%

%\cite{Mulders:1995dh}
\bibitem{MulTan96}
P.~J.~Mulders and R.~D.~Tangerman,
%``The complete tree-level result up to order 1/Q for polarized  deep-inelastic
%leptoproduction,''
Nucl.\ Phys.\ B {\bf 461}, 197 (1996) [Erratum-ibid.\ B {\bf 484},
538 (1997)] [arXiv:hep-ph/9510301].
%%CITATION = HEP-PH 9510301;%%

%\cite{Boer:1997nt}
\bibitem{BoeMul98}
D.~Boer and P.~J.~Mulders,
%``Time-reversal odd distribution functions in leptoproduction,''
Phys.\ Rev.\ D {\bf 57}, 5780 (1998) [arXiv:hep-ph/9711485].
%%CITATION = HEP-PH 9711485;%%%\cite{Boer:2001he}

%\cite{Itzykson:rh}
\bibitem{ItzZub}
C.~Itzykson and J.~B.~Zuber, {\it Quantum Field Theory},
(Mcgraw-Hill, New York, 1980).
%\href{http://www.slac.stanford.edu/spires/find/hep/www?irn=787043}{SPIRES entry}

%\cite{Ji:2002xn}
\bibitem{JiMaYu03}
X.~Ji, J.~P.~Ma and F.~Yuan,
%``Three-quark light-cone amplitudes of the proton and quark-orbital-motion
%dependent observables,''
Nucl.\ Phys.\ B {\bf 652}, 383 (2003) [arXiv:hep-ph/0210430].
%%CITATION = HEP-PH 0210430;%%

%\cite{Jaffe:1991kp}
\bibitem{JafJi91}
R.~L.~Jaffe and X.~D.~Ji,
%``Chiral odd parton distributions and polarized Drell-Yan,''
Phys.\ Rev.\ Lett.\  {\bf 67}, 552 (1991);
%%CITATION = PRLTA,67,552;%%
%\cite{Jaffe:1991ra}
%\bibitem{Jaffe:1991ra}
%R.~L.~Jaffe and X.~D.~Ji,
%``Chiral odd parton distributions and Drell-Yan processes,''
Nucl.\ Phys.\ B {\bf 375}, 527 (1992).
%%CITATION = NUPHA,B375,527;%%

\bibitem{Col93}
%\cite{Collins:1992kk}
%\bibitem{Collins:1992kk}
J.~C.~Collins,
%``Fragmentation of transversely polarized quarks probed in transverse  momentum
%distributions,''
Nucl.\ Phys.\ B {\bf 396}, 161 (1993) [arXiv:hep-ph/9208213].
%%CITATION = HEP-PH 9208213;%%

\bibitem{BroHwaSch02}
%\cite{Brodsky:2002cx}
%\bibitem{Brodsky:2002cx}
S.~J.~Brodsky, D.~S.~Hwang and I.~Schmidt,
%``Final-state interactions and single-spin asymmetries in semi-inclusive  deep
%inelastic scattering,''
Phys.\ Lett.\ B {\bf 530}, 99 (2002) [arXiv:hep-ph/0201296].
%%CITATION = HEP-PH 0201296;%%

%\cite{Qiu:xy}
%\cite{Qiu:xx}
\bibitem{QiuSte91}
J.~w.~Qiu and G.~Sterman,
%``Power Corrections In Hadronic Scattering. 1. Leading 1/Q**2 Corrections To
%The Drell-Yan Cross-Section,''
Nucl.\ Phys.\ B {\bf 353}, 105 (1991);
%%CITATION = NUPHA,B353,105;%%
%\bibitem{Qiu:xy}
%J.~w.~Qiu and G.~Sterman,
%``Power Corrections To Hadronic Scattering. 2. Factorization,''
Nucl.\ Phys.\ B {\bf 353}, 137 (1991).
%%CITATION = NUPHA,B353,137;%%


%\cite{Koike:2003zc}
\bibitem{KoiNag03}
Y.~Koike and J.~Nagashima,
%``Double spin asymmetries for large-p(T) hadron production in  semi-inclusive
%DIS,''
Nucl.\ Phys.\ B {\bf 660}, 269 (2003) [arXiv:hep-ph/0302061].
%%CITATION = HEP-PH 0302061;%%

%\cite{Weber:1991wd}
\bibitem{Web92}
A.~Weber,
%``Soft gluon resummations for polarized Drell-Yan dimuon production,''
Nucl.\ Phys.\ B {\bf 382}, 63 (1992);
%%CITATION = NUPHA,B382,63;%%
%\cite{Weber:1993xm}
%\bibitem{Weber:1993xm}
%A.~Weber,
%``Soft gluon resummation for the single spin production of W+- bosons,''
Nucl.\ Phys.\ B {\bf 403}, 545 (1993).
%%CITATION = NUPHA,B403,545;%%

%\cite{Nadolsky:2003fz}
\bibitem{NadYua03}
P.~M.~Nadolsky and C.~P.~Yuan,
%``Soft parton radiation in polarized vector boson production: Theoretical
%issues,''
Nucl.\ Phys.\ B {\bf 666}, 3 (2003) [arXiv:hep-ph/0304001].
%%CITATION = HEP-PH 0304001;%%

\end{thebibliography}
\end{document}